\documentclass[journal=jpclcd, manuscript=letter, layout=traditional]{achemso}
 \setkeys{acs}{articletitle = true}

\usepackage{wrapfig}
\usepackage{calc}
\usepackage{graphicx}
\usepackage[bookmarks,bookmarksopen]{hyperref}
\usepackage{hyperref}
\usepackage{dcolumn}
\usepackage{bm}
\usepackage[english]{babel}
\usepackage{amsfonts}
\usepackage{amsmath,amssymb}
\usepackage{booktabs}
\usepackage{array}
\usepackage{dcolumn}
\usepackage{setspace}
\usepackage{soul}
\usepackage{color}
\usepackage{setspace}
\usepackage{threeparttable}
\usepackage{multirow}
\usepackage{cleveref}
\usepackage{amssymb}
\usepackage{float}
\usepackage{subfloat}
\usepackage{subfig}
\usepackage{graphicx}
\usepackage{tabularx, booktabs} 
\newcolumntype{Y}{>{\centering\arraybackslash}X}
\usepackage[width=.85\textwidth]{caption}

\newcolumntype{L}[1]{>{\raggedright\let\newline\\\arraybackslash\hspace{0pt}}m{#1}}
\newcolumntype{C}[1]{>{\centering\let\newline\\\arraybackslash\hspace{0pt}}m{#1}}
\newcolumntype{R}[1]{>{\raggedleft\let\newline\\\arraybackslash\hspace{0pt}}m{#1}}
\usepackage{longtable}

\newcommand{\be}{\begin{equation}}
\newcommand{\ee}{\end{equation}}

\newcommand{\cm}{cm$^{-1}$}
\newcommand{\cms}{cm$^{-1}$ }

\title{ 
	Direct signatures of light-induced conical intersections on the field-dressed spectrum of Na$_{2}$} 
	
	\author{Tam\'as Szidarovszky}
	
	\email{tamas821@caesar.elte.hu}
	
	\affiliation{Laboratory of Molecular Structure and Dynamics, Institute of Chemistry,
	E\"otv\"os Lor\'and University and MTA-ELTE Complex Chemical Systems Research Group,
	H-1117 Budapest, P\'azm\'any P\'eter s\'et\'any 1/A, Hungary}
	
	\author{G\'abor J. Hal\'asz}
	
	\affiliation{Department of Information Technology, University of Debrecen, PO Box 400, H-4002 Debrecen, Hungary}

	\author{Attila G. Cs\'asz\'ar}
	
	\affiliation{Laboratory of Molecular Structure and Dynamics, Institute of Chemistry,
	E\"otv\"os University and MTA-ELTE Complex Chemical Systems Research Group,
	H-1117 Budapest, P\'azm\'any P\'eter s\'et\'any 1/A, Hungary}
	
	\author{Lorenz S. Cederbaum}
	
	\affiliation{Theoretische Chemie, Physikalisch-Chemisches Institut, Universität Heidelberg, D-69120, Heidelberg, Germany}	
	
	\author{\'Agnes Vib\'ok}
	
	\email{vibok@phys.unideb.hu}
	
	\affiliation{Department of Theoretical Physics, University of Debrecen, PO Box 400, H-4002 Debrecen, Hungary and ELI-ALPS, ELI-HU Non-Profit Ltd., Dugonics t\'er 13, H-6720 Szeged, Hungary}	
	
	\date{\today}

	\date{\today}

\begin{document}

\begin{abstract}
Rovibronic spectra of the field-dressed homonuclear diatomic Na$_2$ molecule is investigated to identify direct signatures of the light-induced conical intersection (LICI) on the spectrum. 
The theoretical framework formulated allows the computation of the (1) field-dressed rovibronic states induced by a medium intensity continuous-wave laser light and the (2) transition amplitudes between these field-dressed states with respect to an additional weak probe pulse.
The field-dressed spectrum features absorption peaks resembling the field-free spectrum as well as stimulated emission peaks corresponding to transitions not visible in the field-free case.
By investigating the dependence of the field-dressed spectra on the dressing-field wavelength, both in full- and reduced dimensional simulations, direct signatures of the LICI can be identified.
These signatures include (1) the appearance of new peaks and the splitting of peaks for both absorption and stimulated emission, and (2) the manifestation of an intensity borrowing effect in the field-dressed spectrum.
\end{abstract}
\maketitle

\section{Introduction}


Theoretical and experimental studies have revealed numerous new
phenomena resulting from the interaction of matter with strong laser fields, such as high harmonic generation,
\cite{HHG,Corkum} above threshold ionization, \cite{Kulander} dissociation,
\cite{Bandrauk1} and bond softening and hardening effects.\cite{Bandrauk2,Bandrauk3,Bucksbaum1,Bucksbaum2}
The last two of these phenomena can be explained by adopting the dressed-state
or light-induced potential (LIP) picture, which involves a nuclear degree of freedom (most commonly the molecular vibration)
in addition to the electronic degrees of freedom in the case of molecules. \cite{Wunderlich1,Figger1,Sola_LIP} 
LIP provides an appropriate interpretation of the dissociation processes of small
molecules exposed to high-intensity laser field. For example, LIP
predicts that at low light intensities the dissociation rate of molecules behaves
linearly with intensity according to the Fermi's golden rule (FGR),
while at larger intensities the dissociation rate is rather strongly nonlinear.\cite{Wunderlich1} 

Including a second nuclear degree of freedom in the Hamiltonian
of the light-matter interaction results in the ``light-induced
conical intersection'' (LICI) picture. LICI was first discussed for
diatomics under the presence of a standing external electric field. \cite{Nimrod1}
In this situation the translational motion of the molecule can strongly
couple with the electronic degrees of freedom, while the vibrational motion
provides the second dynamical variable, forming a branching
space in diatomics and leading to a periodic array of light induced
CIs. However, when a running laser field is present, the vibrational and the rotational
degrees of freedom serve as the two dynamical variables.\cite{Nimrod2,LICI1,LICI2} Recently, a number of studies appeared about the nature of the LICIs. Numerous theoretical and experimental
studies have confirmed that LICIs have noticeable impact on different
dynamical properties (like molecular alignment, photodissociation
probability, etc...). \cite{LICI1,LICI2,LICI3,LICI4,LICI5,Bandrauk_LICI} Furthermore, a strong
effect in the angular distribution of the photofragments has been
revealed that serves as a direct signature of the LICI.\cite{LICI5}
The first experimental observation of LICIs in diatomic molecules
was made by Bucksbaum \textit{et al.}.\cite{LICI6} Besides
the diatomic studies, a few results are available for polyatomics
as well.\cite{Martinez_LICI,LICI7,LICI8,LICI9} In this case, due to the presence
of several vibrational degrees of freedom, LICI can form without rotation,
which opens up the door for manipulating and controlling nonadiabatic
effects by light.

Previous studies mainly concentrated on the dynamics of diatomic systems under the influence of strong electric fields.
In the present work, we focus on the description
of the field-dressed static rovibronic spectrum of diatomics. The
theoretical and experimental investigations of the field-free rovibronic
spectrum of diatomic molecules have been studied extensively for more than a
century,\cite{spectroscopy_Herzberg,spectroscopy_HHRS}
significantly contributing to our fundamental understanding of chemical
and physical phenomena. Measuring or computing spectral
transitions between field-dressed states is well developed for atoms,\cite{CohenTannoudji}
and to some extent has also been incorporated to molecular spectroscopy.
For example, inducing Autler--Townes-type splittings\cite{AutlerTownes_original} of 
rotational transitions with microwave radiation have been used to promote the assignation
of rovibrational spectra \cite{AutlerTownes_in_spectroscopy_1997} or to deduce molecular parameters, such as the transition dipole moment.\cite{AutlerTownes_in_spectroscopy_2017,AutlerTownes_in_spectroscopy_2012}
Theoretical work considering the transitions between field-dressed rovibronic states was also done for a dressing field of specific wavelength.\cite{Nimrod2}

The purpose of the present work is to investigate the frequency- and intensity dependence of the field-dressed
rovibronic spectrum of the homonuclear diatomic molecule Na$_{2}$
and to identify the direct signatures of LICIs on the spectrum.

\section{Theoretical approach}

\label{Theory}

In this study we are providing a framework to simulate the weak-field absorption and stimulated emission spectrum
of field-dressed homonuclear diatomic molecules. First, we determine
the field-dressed states, which we assume are formed from the field-free
eigenstates of the system through the interaction with an external
laser field of medium intensity, turned on adiabatically
with respect to the characteristic molecular timescales.
Second, we
compute the dipole transition amplitudes between the field-dressed
states with respect to a weak second pulse, using first order time-dependent
perturbation theory.\cite{LANDAU_QM} The showcase example is the
Na$_{2}$ molecule.


\subsection{The field-dressed Floquet Hamiltonian of Na$_2$}

Using the dipole approximation and considering only the ground $V_{1}(R)={\rm X}^{1}\Sigma{\rm _{g}^{+}}$ and the first excited $V{}_{2}(R)={\rm A}^{1}\Sigma{\rm _{u}^{+}}$ electronic states of Na$_2$, the time-dependent (TD) Hamiltonian of the system can be represented in the Floquet picture \cite{Floquet} and can be written in a block diagonal from after neglecting the off-resonant light-matter coupling terms. The $N$th block of the Hamiltonian reads
\begin{equation}
\hat{H}(N)=\left[-\frac{\hbar^{2}}{2\mu}\frac{\partial^{2}}{\partial R^{2}}+
\frac{\mathbf{L}_{\theta\varphi}^{2}}{2\mu R^{2}}\right]
\otimes
\begin{bmatrix}
1 & 0 \\ 0 & 1
\end{bmatrix}
+\begin{bmatrix}V_{1}(R)+N\hslash\omega & (F_{0}/2)d(R){\rm cos}(\theta)\\
(F_{0}/2)d(R){\rm cos}(\theta) & V_{2}(R)+(N-1)\hslash\omega
\end{bmatrix},\label{eq:2by2Hamiltonian}
\end{equation}
where, $R$ and ($\theta,\varphi$) are the molecular vibrational and
rotational coordinates, respectively, $\mu$ denotes the reduced mass
of the diatomic, and $\mathbf{L}_{\theta\varphi}$ is the rotational angular
momentum operator of the nuclei. $F_{0}$ and $\omega$ are
the amplitude and the frequency of the dressing electric
field, respectively.
 $d(R)$ is the dipole transition matrix element in the body-fixed frame and $\theta$ also describes the angle
between the polarization direction of the linearly polarized electric
field and the direction of the transition dipole ($\hslash$ is Planck's constant divided
by $2\pi$).

After diagonalizing the diabatic potential matrix of Eq. (\ref{eq:2by2Hamiltonian}), the resulting
adiabatic or light-induced surfaces form a conical intersection for geometry parameters
determined by the following conditions: $V_{1}(R)=V_{2}(R)-\hslash\omega$
and $\theta=\pi/2$.\cite{LICI5}

One can obtain all field-dressed (FD) eigenstates $\vert\Psi_{i}^{{\rm FD}}(N)\rangle$
and quasi-energies $E_{i}^{{\rm FD}}(N)$ by determining the eigenpairs
of the Hamiltonian of Eq. (\ref{eq:2by2Hamiltonian}) and applying
an energy shift of $k\hslash\omega$ to the quasienergies as necessary,
where $k$ is an integer. 

We diagonalize the Hamiltonian of Eq. (\ref{eq:2by2Hamiltonian})
after constructing its matrix representation using the basis of field-free
rovibrational states.
The field-free eigenstates of Na$_{2}$, \textit{i.e.}, the basis functions
used to expand the field-dressed states, can be characterized by three
quantum numbers and are represented as $\vert jvJ\rangle$, where
the molecule is in the $j$th electronic, $v$th vibrational, and
$J$th rotational state, and $j=1$ and $j=2$ stand for the ${\rm X}^{1}\Sigma{\rm _{g}^{+}}$
and ${\rm A}^{1}\Sigma{\rm _{u}^{+}}$ electronic states, respectively.

Using the notation termed "\textit{Floquet-state nomenclature}" in Ref. \citenum{Floquet}, 
the field-dressed states can be expressed as the linear combination
of products of field-free molecular rovibronic states and the Fourier
vectors of the Floquet states, $i.e.$, 
\begin{equation}
\vert\Psi_{i}^{{\rm FD}}(N)\rangle=\sum_{J,v}C_{i,1vJ}\vert1vJ\rangle\vert N\rangle+\sum_{J,v}C_{i,2vJ}\vert2vJ\rangle\vert N-1\rangle\label{eq:FieldDressedStates_large_photon_number}
\end{equation}
where $\vert N\rangle$ is the $N$th Fourier vector of the Floquet state, and $C_{i,jvJ}$ are the expansion coefficients
obtained by diagonalizing the Hamiltonian of Eq. (\ref{eq:2by2Hamiltonian})
in the basis of the field-free rovibrational states.
As detailed in Ref. \citenum{Floquet}, the time-dependence of the periodic Floquet state is determined by the time-dependence of its Fourier vectors $\langle t \vert N\rangle = e^{iN \omega t}$.

\subsection{Transitions between field-dressed states}

Given the Hamiltonian and the field-dressed states, 
we now turn to the computation of the spectrum of field-dressed molecules, \textit{i.e.}, we turn to the computation of transition amplitudes between field-dressed states as induced by a weak probe pulse. Since we assume the probe pulse to be weak, transitions induced by it should be dominated by one-photon processes. Therefore, following the standard approach of theoretical molecular spectroscopy,\cite{BunkerJensen} we use first-order TD perturbation theory to compute the transition 
amplitudes induced by the probe pulse. Then, the amplitudes are proportional to 
$\langle \Psi_i^{\rm FD} \vert \bold{\hat{d}\hat{e}} \vert \Psi_j^{\rm FD} \rangle  = \langle\Psi_{i}^{{\rm FD}}\vert\hat{d}{\rm cos}(\theta)\vert\Psi_{j}^{{\rm FD}}\rangle$, where $\bold{\hat{e}}$ is a unit vector defining the polarization direction of the probe pulse, which we assume to be identical to that of the dressing pulse.
The conservation of energy requires that $E_{j}^{{\rm FD}}=E_{i}^{{\rm FD}}\pm\hbar\omega_{\rm p}$,
where $\omega_{\rm p}$ is the angular frequency of the weak probe pulse.
The matrix element of the operator $\hat{d}{\rm cos}(\theta)$ between
two field-dressed states of Eq. (\ref{eq:FieldDressedStates_large_photon_number})
gives \begin{footnotesize} 
\begin{equation}
\begin{aligned} & \langle\Psi_{i}^{{\rm FD}}(N)\vert\hat{d}{\rm cos}(\theta)\vert\Psi_{j}^{{\rm FD}}(N')\rangle=\\
 & =\left(\sum_{J,v}C_{i,1vJ}^{*}\langle1vJ\vert\langle N\vert+\sum_{J,v}C_{i,2vJ}^{*}\langle2vJ\vert\langle N-1\vert\right)\hat{d}{\rm cos}(\theta)\left(\sum_{J',v'}C_{j,1v'J'}\vert1v'J'\rangle\vert N'\rangle+\sum_{J',v'}C_{j,2v'J'}\vert2v'J'\rangle\vert N'-1\rangle\right)=\\
 & =\sum_{J,v,J',v'}C_{i,1vJ}^{*}C_{j,2v'J'}\langle1vJ\vert\hat{d}{\rm cos}(\theta)\vert2v'J'\rangle\langle N\vert N'-1\rangle+\sum_{J,v,J',v'}C_{i,2vJ}^{*}C_{j,1v'J'}\langle2vJ\vert\hat{d}{\rm cos}(\theta)\vert1v'J'\rangle\langle N-1\vert N'\rangle=\\
 & =\sum_{J,v,J',v'}C_{i,1vJ}^{*}C_{j,2v'J'}\langle1vJ\vert\hat{d}{\rm cos}(\theta)\vert2v'J'\rangle\delta_{N,N'-1}+\sum_{J,v,J',v'}C_{i,2vJ}^{*}C_{j,1v'J'}\langle2vJ\vert\hat{d}{\rm cos}(\theta)\vert1v'J'\rangle\delta_{N,N'+1}.
\end{aligned}
\label{eq:transition_amplitude_between_classical_FD_states}
\end{equation}
\end{footnotesize} 

In Eq. (\ref{eq:transition_amplitude_between_classical_FD_states})
we exploited the fact that Na$_2$ has no permanent dipole, and that the $\vert N\rangle$ Fourier vectors
are orthogonal with respect to the scalar product defined in the time-domain of the extended Hilbert space of Floquet theory.\cite{Floquet}

In the last line of Eq. (\ref{eq:transition_amplitude_between_classical_FD_states}),
the first term represents transitions with transition frequencies
$\nu_{ij}=(hc)^{-1}\vert E_{j}^{{\rm FD}}(N)+\hslash\omega-E_{i}^{{\rm FD}}(N)\vert$,
to which the first electronic state contributes through the $i$th
field-dressed state and the second electronic state contributes through
the $j$th field-dressed state. 

As will be shown, the first term in
the last line of Eq. (\ref{eq:transition_amplitude_between_classical_FD_states})
leads to the usual field-free absorption spectrum in the limit of
the dressing field intensity going to zero. On the other hand, the
second term in the last line of Eq. (\ref{eq:transition_amplitude_between_classical_FD_states})
represents transitions with transition frequencies $\nu_{ij}=(hc)^{-1}\vert E_{j}^{{\rm FD}}(N)-\hslash\omega-E_{i}^{{\rm FD}}(N)\vert$,
in which the second electronic state contributes through the $i$th
field-dressed state and the first electronic state contributes through
the $j$th field-dressed state. For all dressing-field wavelengths
considered in this study, the second term in the last line of Eq.(\ref{eq:transition_amplitude_between_classical_FD_states})
represents stimulated emission from the $\vert\Psi_{i}^{{\rm FD}}(N)\rangle$
state, as induced by the probe pulse.

\section{Results and discussion}

\begin{figure}[H]
\begin{centering}
\includegraphics[height=75mm]{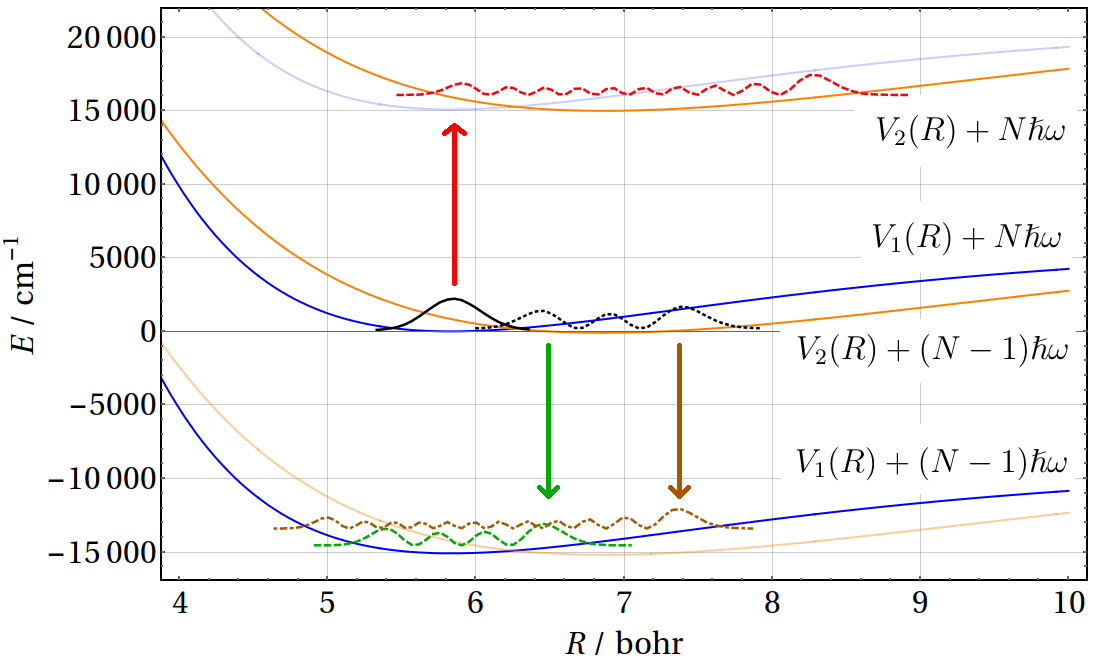} 
\par\end{centering}

\caption{Field-dressed diabatic PECs of Na$_{2}$ obtained with a dressing-light wavelength of
$\lambda=657$ nm. The energy scale stands for quasienergy. 
Vibrational probability densities are drawn for the $\vert1$ $0$ $0\rangle\vert N \rangle$ (continous black line on the $V_{1}(R)+N\hslash\omega$ PEC), $\vert1$ $3$ $0\rangle\vert N-1 \rangle$ (green dashed line on the $V_{1}(R)+(N-1)\hslash\omega$ PEC), $\vert1$ $11$ $0\rangle\vert N-1 \rangle$ (brown dashed line on the $V_{1}(R)+(N-1)\hslash\omega$ PEC), $\vert2$ $2$ $1\rangle\vert N-1 \rangle$ (black dotted line on the $V_{2}(R)+(N-1)\hslash\omega$ PEC), and $\vert2$ $9$ $1\rangle\vert N \rangle$ (red dashed line on the $V_{2}(R)+N\hslash\omega$ PEC) states. 
Upward- and downward pointing vertical arrows represent
transitions of absorption and stimulated emission, respectively. The two product states with the largest contribution
to the field-dressed state correlating to $\vert1$ $0$ $0\rangle$ at $F_0\rightarrow0$
are $\vert1$ $0$ $0\rangle\vert N \rangle$ and $\vert2$ $2$ $1\rangle\vert N-1 \rangle$.
}
\label{Fig:Fig1} 
\end{figure}

Here we compute the field-dressed states and related spectra of the Na$_2$ molecule.
Figure \ref{Fig:Fig1} demonstrates the field-dressed diabatic potential energy curves (PEC) of the 
${\rm X}^{1}\Sigma{\rm _{g}^{+}}$ and ${\rm A}^{1}\Sigma{\rm _{u}^{+}}$
electronic states of Na$_{2}$ as well as the vibrational probability densities
of the $\vert1$ $0$ $0\rangle\vert N \rangle$, $\vert1$ $3$ $0\rangle\vert N-1 \rangle$, $\vert1$ $11$ $0\rangle\vert N-1 \rangle$, $\vert2$ $2$ $1\rangle\vert N-1 \rangle$, and $\vert2$ $9$ $1\rangle\vert N \rangle$ states.

\subsection{Field-dressed spectra and their interpretation}

In the following, we consider spectra obtained from a single $\vert\Psi_{i}^{{\rm FD}}(N)\rangle$
field-dressed state, which we imagine to be populated adiabatically
from the field-free ground state, $\vert1$ $0$ $0\rangle$, by slowly turning
on the dressing field. In all spectra shown, we plot the absolute
square of transition amplitudes as computed by Eq. (\ref{eq:transition_amplitude_between_classical_FD_states}),
or their convolution with a Gaussian function having $\sigma = 50$ cm$^{-1}$.

\begin{figure}[H]
\begin{centering}
\includegraphics[width=0.45\textwidth]{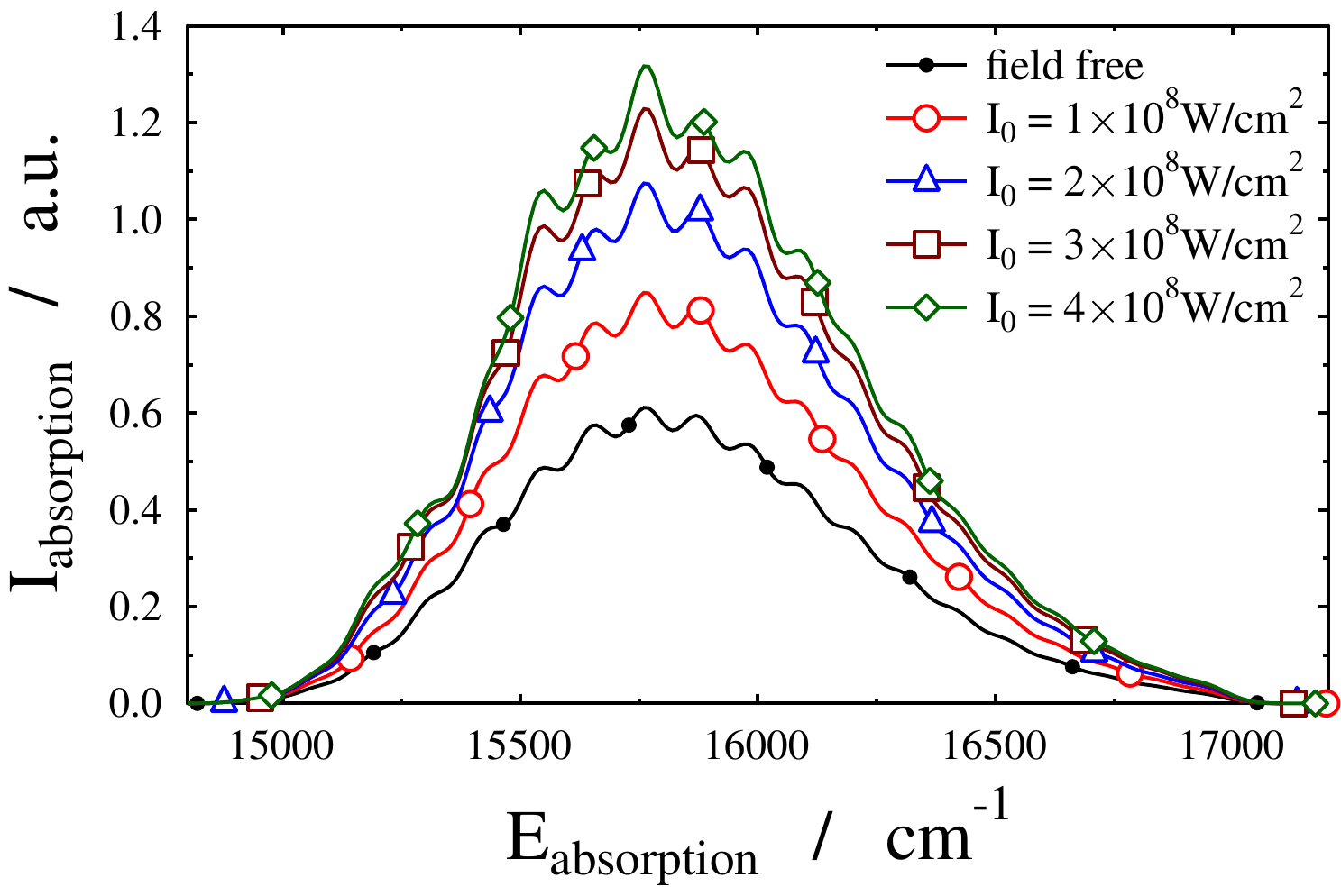}
\includegraphics[width=0.45\textwidth]{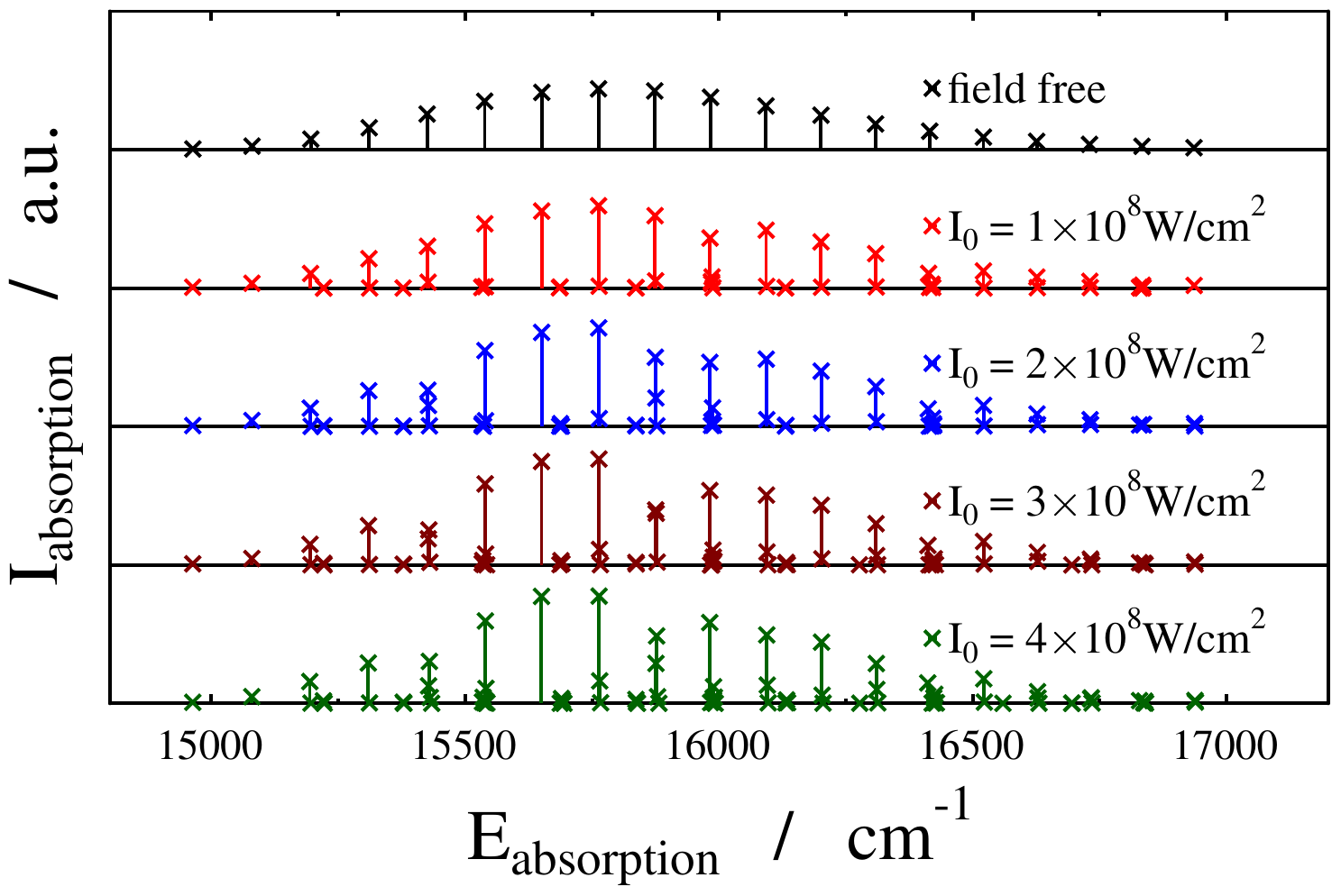} 
\includegraphics[width=0.45\textwidth]{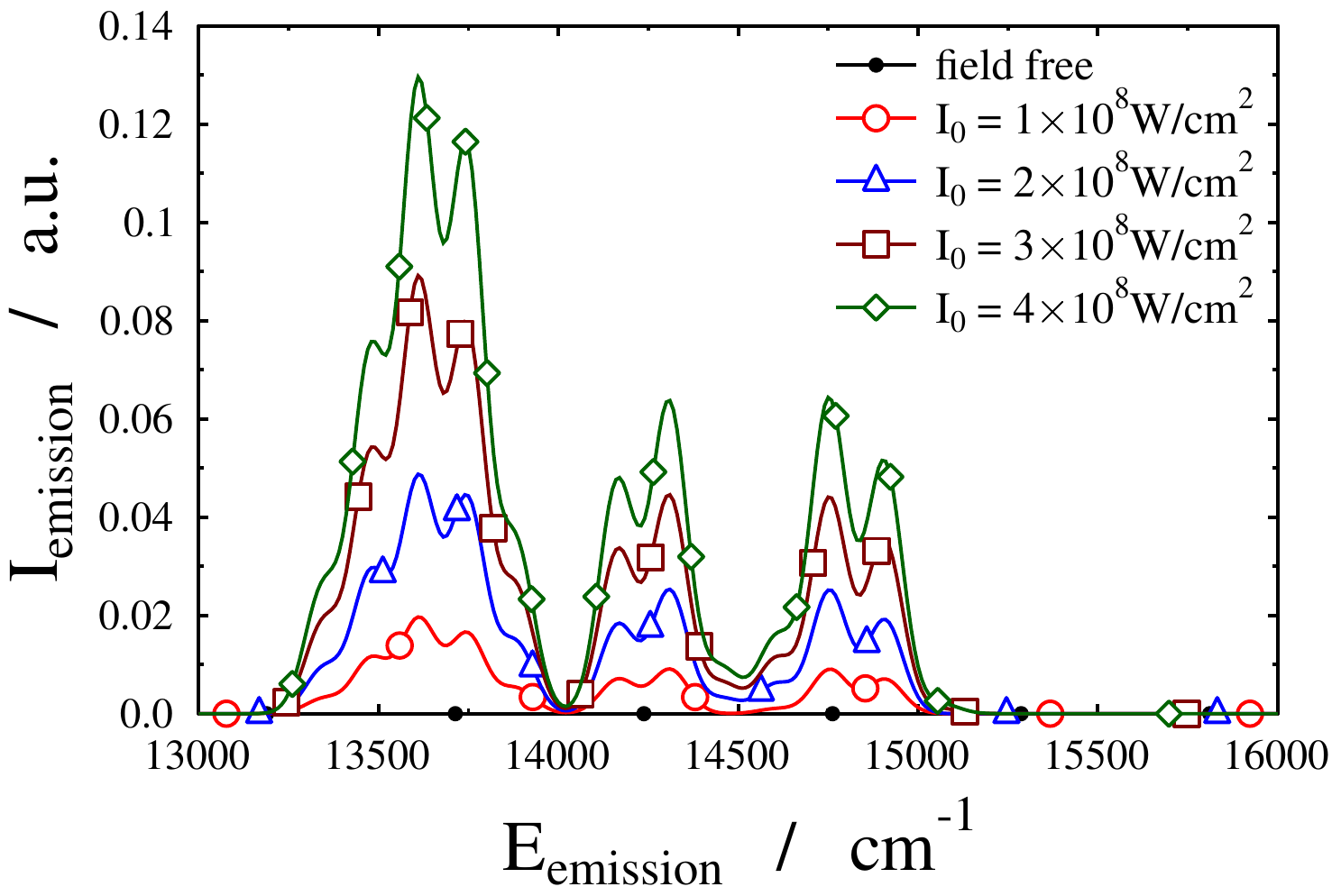} 
\includegraphics[width=0.45\textwidth]{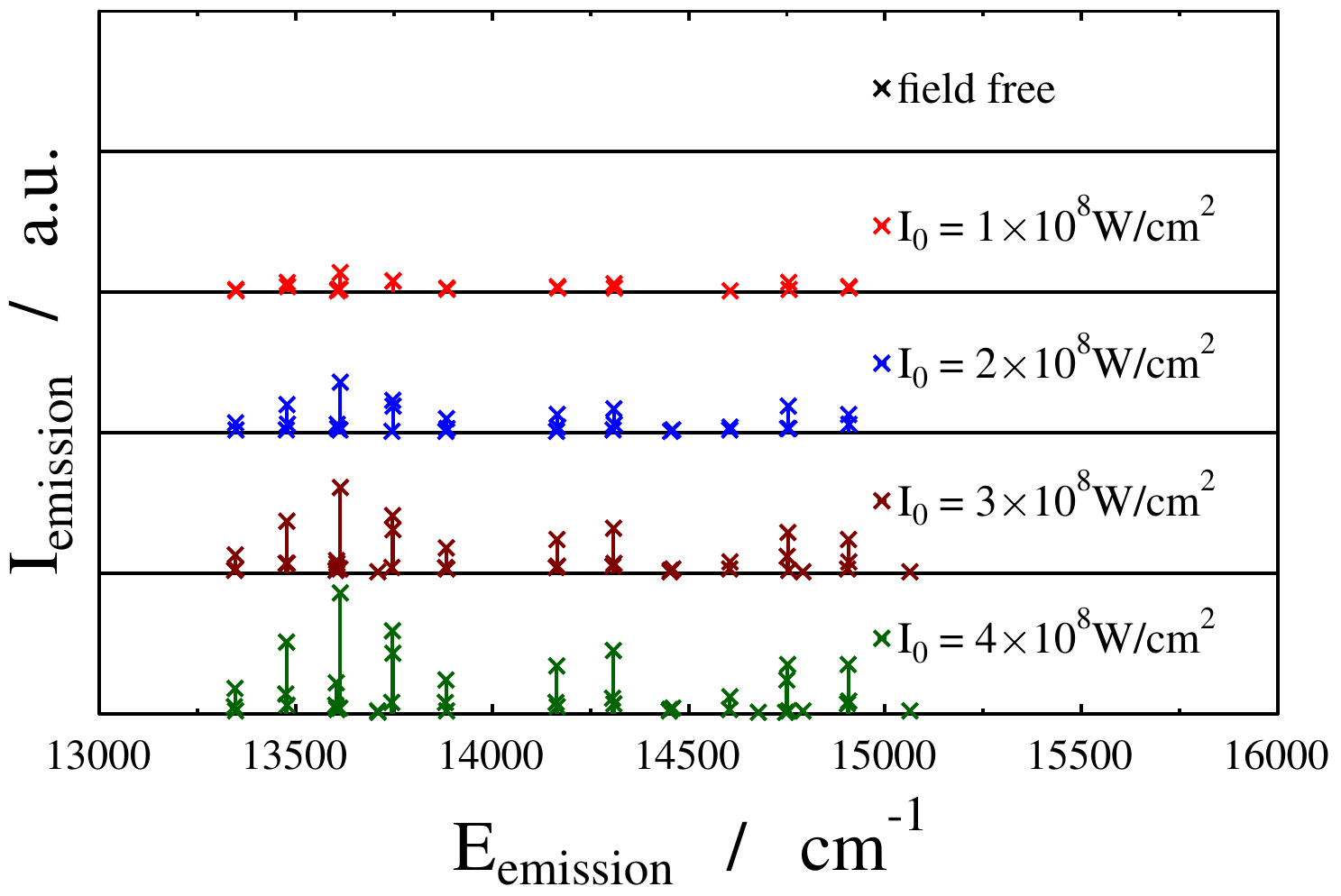} 
\par\end{centering}

\caption{One-photon spectra of the field-dressed Na$_{2}$ as a function of
dressing-field ($\lambda=657$ nm) intensity, computed using Eq. \ref{eq:transition_amplitude_between_classical_FD_states}.
The spectra only show transitions from the field-dressed state which
becomes the rovibrational ground state $\vert1$ $0$ $0\rangle$ for $F_{0}\rightarrow0$. The upper two panels show absorption peaks, while the lower two panels present stimulated emission peaks. The curves shown in the left panels are obtained by convolving the right hand side spectra with a Gaussian function having $\sigma=50$ cm$^{-1}$.}
\label{Fig:Fig2} 
\end{figure}

Figure \ref{Fig:Fig2} shows field-dressed
spectra of Na$_{2}$ obtained with a dressing-light wavelength of
$\lambda=657$ nm, which is nearly resonant with the $\vert1$ $0$ $0\rangle\leftrightarrow\vert2$ $2$ $1\rangle$
transition. In this case the two product states with the largest contribution
to the field-dressed state correlating to $\vert1$ $0$ $0\rangle$ at $F_0\rightarrow0$
are $\vert1$ $0$ $0\rangle\vert N \rangle$ and $\vert2$ $2$ $1\rangle\vert N-1 \rangle$, with $\vert1$ $0$ $J\rangle\vert N \rangle$ (even $J$) and $\vert 2$ $2$ $J \pm 1 \rangle\vert N-1 \rangle$  states giving further contributions, whose magnitudes depend on the dressing-field intensity. Figure \ref{Fig:Fig2} shows that
for low dressing-field intensities one basically obtains the field-free
absorption spectrum of Na$_{2}$ with all peaks above $15000$ \cm
, resulting from transitions between the $\vert1$ $0$ $0\rangle$ ground
state and the  $\vert2$ $v$ $1\rangle$ rovibronic states. Due to the significantly
different equilibrium distances of the ${\rm X}^{1}\Sigma{\rm _{g}^{+}}$
and ${\rm A}^{1}\Sigma{\rm _{u}^{+}}$ PECs, see Figure \ref{Fig:Fig1}, the $\vert1$ $0$ $0\rangle$
ground state has considerable Frank--Condon (FC) overlaps with many vibrational
states of ${\rm A}^{1}\Sigma{\rm _{u}^{+}}$, leading to a dozen or
so absorption peaks even in the field-free case. This group of peaks located above $15000$ \cms originates from the contribution of the
first term in the last line of Eq. (\ref{eq:transition_amplitude_between_classical_FD_states}).
At higher dressing-field intensities the profile of this group of
peaks changes, involving peaks primarily originating from transitions between the $\vert1$ $0$ $J\rangle\vert N \rangle$ low-lying rotational states of ${\rm X}^{1}\Sigma{\rm _{g}^{+}}$ and highly excited rovibrational states on ${\rm A}^{1}\Sigma{\rm _{u}^{+}}$ of the type $\vert 2$ $v$ $J \pm 1 \rangle\vert N \rangle$, where $J$ is even.

Figure \ref{Fig:Fig2} also shows that for
increasing dressing-field intensities another group of transitions
(composed of three subgroups) appears between 13000 and 15500
\cm . The origin of these peaks is the second term in the last line
of Eq. (\ref{eq:transition_amplitude_between_classical_FD_states})
and they show transitions in which the quasienergy of the final state
is lower than that of the initial state, \textit{i.e.}, they are stimulated emission peaks.
Based on Figure \ref{Fig:Fig1}, one can conclude 
that the stimulated emission peaks between $13000$ and $15500$ \cms in
Figure \ref{Fig:Fig2} primarily originate
from transitions between the low-lying $\vert2$ $2$ $J\rangle\vert N-1 \rangle$ rovibrational
states and the vibrationally highly excited $\vert 1$ $v$ $J \pm 1 \rangle\vert N-1 \rangle$ rovibrational
states, where $J$ is odd. The three distinct groups of stimulated emission
peaks seen in Figure \ref{Fig:Fig2} is a clear
indication of the FC overlaps between $\vert2$ $2$ $J\rangle$ and the vibrationally
excited states of $\vert1$ $v$ $J \pm 1\rangle$ type.

\subsection{Signatures of LICIs in the field-dressed spectrum}

\begin{figure}[H]
\begin{centering}
\includegraphics[width=0.45\textwidth]{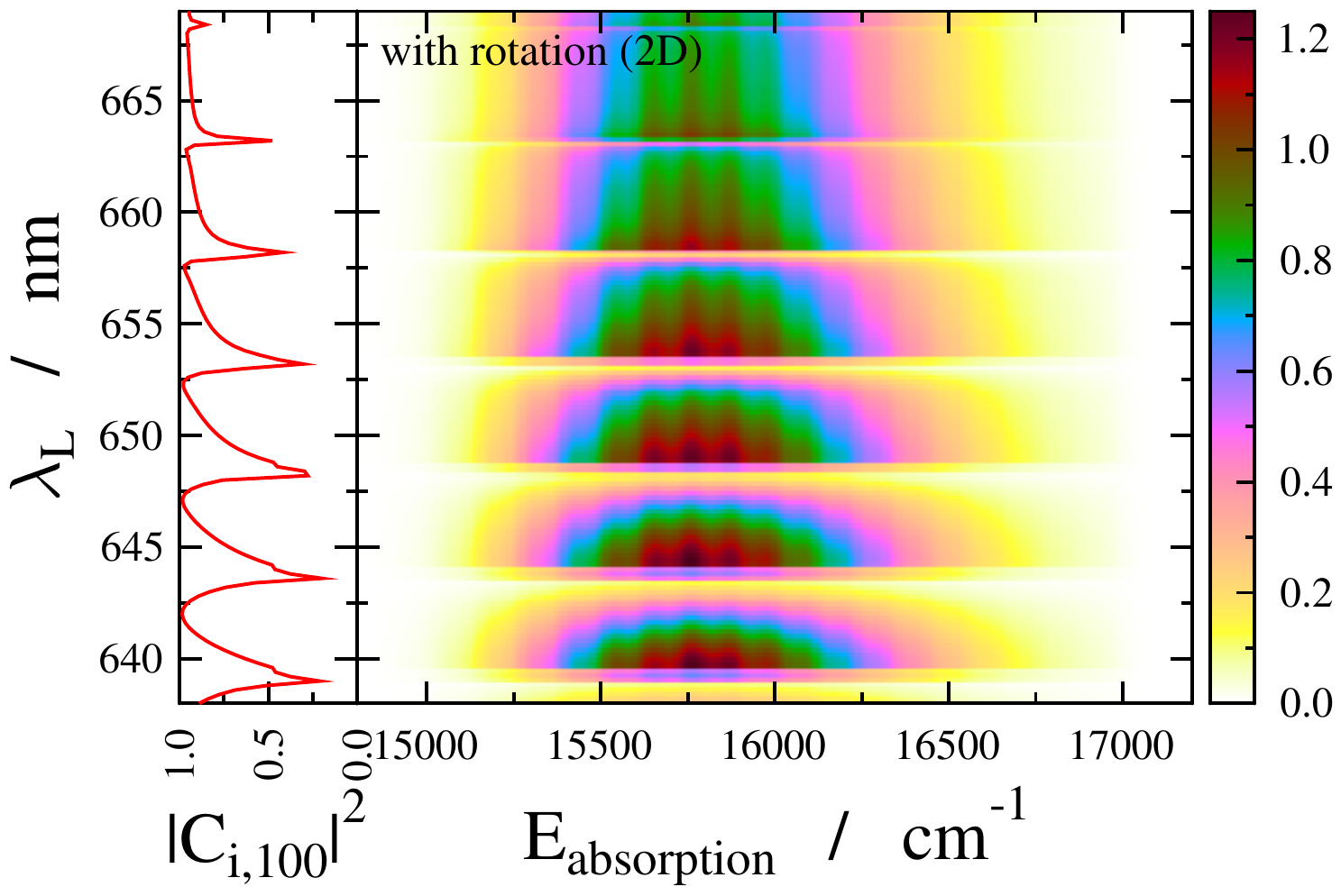} 
\includegraphics[width=0.45\textwidth]{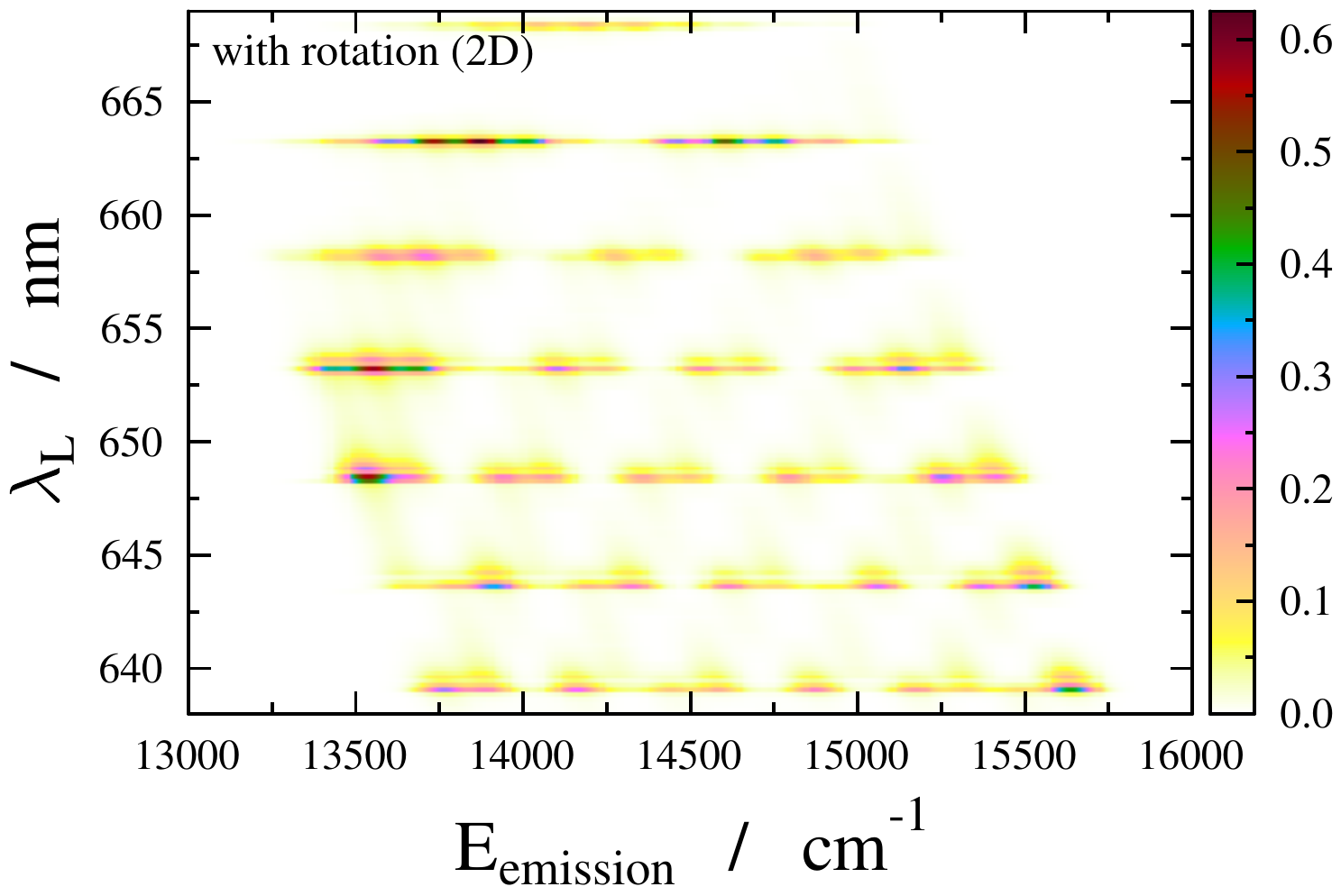} 
\par\end{centering}

\caption{
One-photon spectra of the field-dressed Na$_{2}$ as a function of
dressing-field wavelength at a dressing-field
intensity of $I{_0}=10^{8}$ Wcm$^{-2}$ , computed using Eq. (\ref{eq:transition_amplitude_between_classical_FD_states}) and convolving the spectra at each fixed dressing-field wavelength with a Gaussian function having $\sigma=50$ cm$^{-1}$. The left panel shows absorption, while the right panel presents stimulated emission peaks. The left side of the left panel shows the absolute square of the coefficient of the field-free ground state for the field-dressed state correlating
to $\vert1$ $0$ $0\rangle$ at $I{_0}\rightarrow0$ Wcm$^{-2}$, which was used to obtain the spectra.
} 
\label{Fig:Fig3} 
\end{figure}

By changing the dressing-field frequency from 657 nm, the field-dressed states and their corresponding spectra naturally
change as well, as presented in Figure \ref{Fig:Fig3}. 
From now on we focus on how the presence of a LICI can directly
be observed in field-dressed spectra. In this regard, we point out that, as Figure \ref{Fig:Fig3} shows, the
spectra feature prominent variations in the absorbtion peak intensities as well as significant stimulated emission, when the dressing-field wavelength is near a resonant transition of $\vert1$ $0$ $0\rangle\leftrightarrow\vert2$ $v$ $1\rangle$
type. Thus, as expected, the most significant mixing of field-free
states in the dressed state correlating to $\vert1$ $0$ $0\rangle$ at $F_0\rightarrow0$
occur at the vicinity of resonant frequencies. This strong mixing can lead to the correlation between light-dressed states
and field-free states becoming unambigous. Therefore, in order to avoid any confusion arising from comparing spectra originating from
different field-dressed states, we investigate the effects of a LICI
in light-dressed spectra obtained with nonresonant dressing fields. The wavelength of these nonresonant fields were chosen to be located exactly halfway between resonant wavelengths.

We aim to identify features in the field-dressed spectra which resemble
one of the two following characteristic features of LICIs in diatomics:
(1) coupling between the vibrational and a rotational degree of freedom
manifested in the $\theta$-dependent adiabatic PECs, and (2) nonadiabatic
couplings between the two adiabatic field-dressed elecronic states
connected by the LICI.

\begin{figure}[H]
	\begin{centering}
	\includegraphics[width=0.5\textwidth]{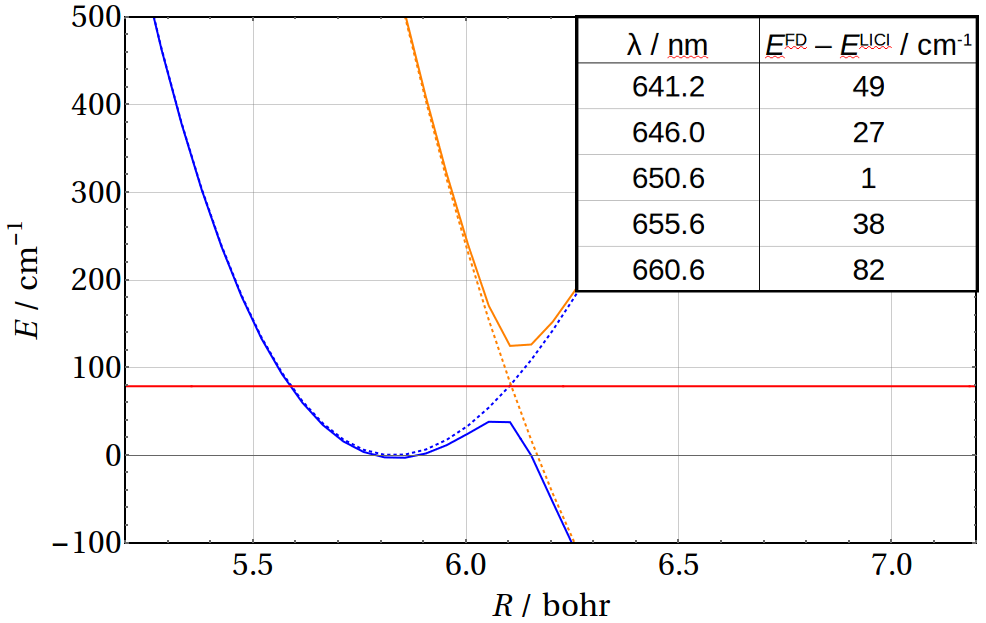} 
	\includegraphics[width=0.5\textwidth]{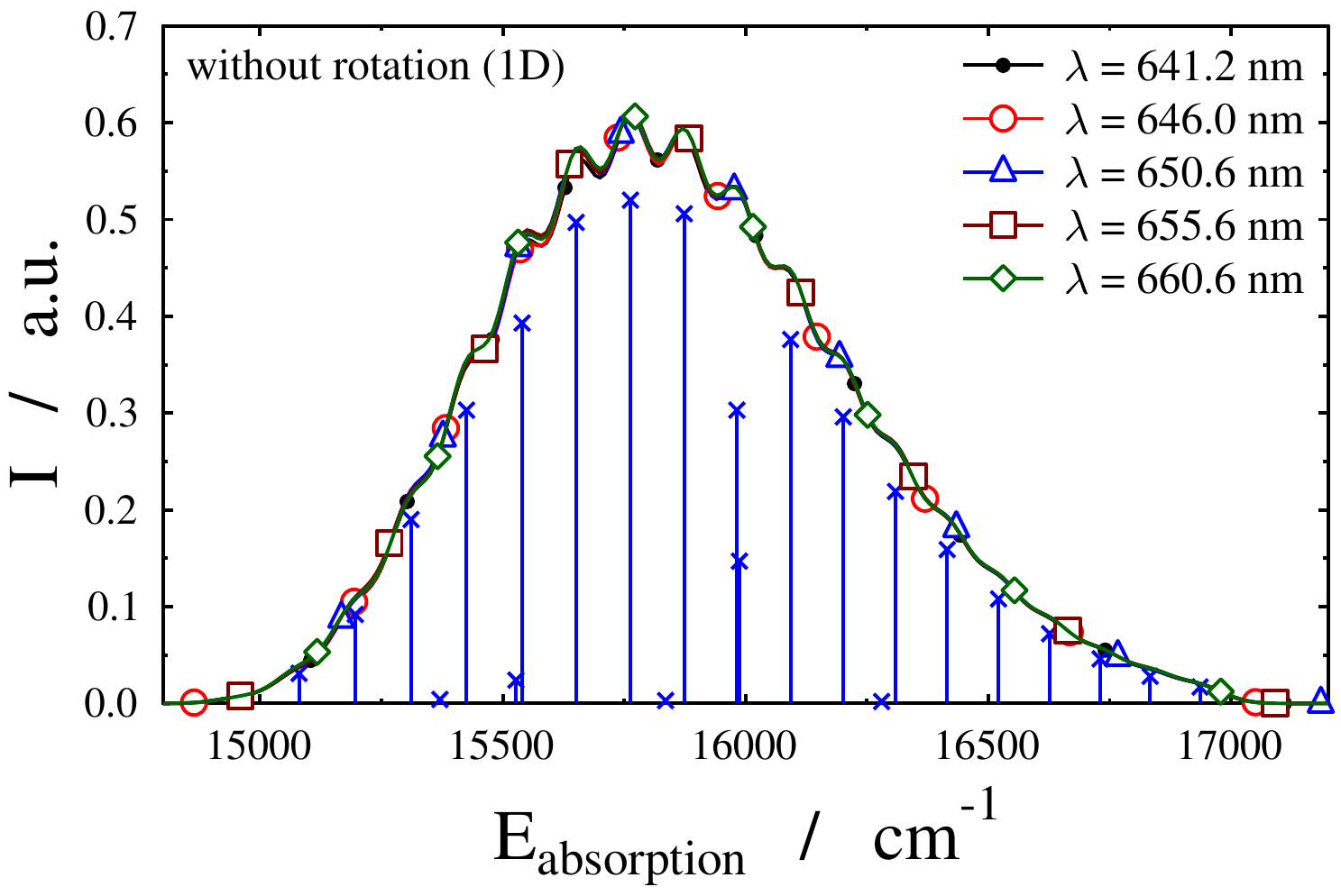} 
	\includegraphics[width=0.5\textwidth]{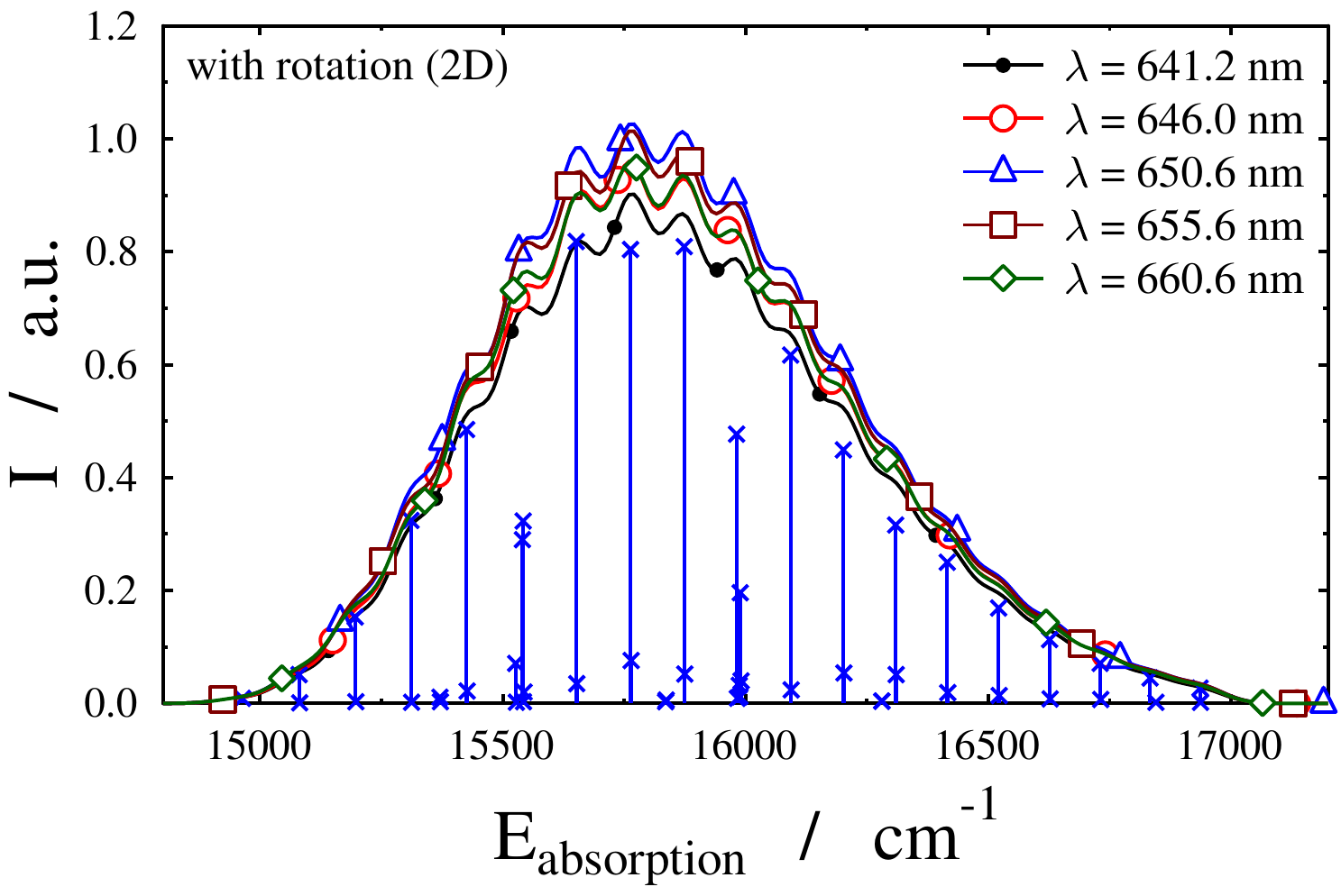} 
	\par\end{centering}

	\caption{Absorption spectra of the field-dressed Na$_{2}$ at five non-resonant
dressing-field wavelengths and a dressing-field intensity of $I=10^{8}$ Wcm$^{-2}$, obtained within the 1D (middle panel) and 2D (lowest panel) model, see text. The stick spectra in the middle and lowest panels represent the spectra obtained with the dressing-field wavelength of $\lambda=650.6$ nm.
In the upmost panel diabatic and adiabatic PECs at $\theta=0$ are shown along with
the quasienergy of the field-dressed state correlating to $\vert1$ $0$ $0\rangle$
at $F_0\rightarrow0$ for $\lambda = 650.6$ nm. The table inset in the upmost panel shows for different dressing-field wavelengths the difference between the energy position of the LICI and the quasienergy of the field-dressed state correlating to $\vert1$ $0$ $0\rangle$
at $F_0\rightarrow0$.}
	\label{Fig:Fig4}
\end{figure}

Figure \ref{Fig:Fig4} shows spectra of field-dressed
Na$_{2}$ obtained for five non-resonant dressing-field wavelengths (641.2, 646.0, 650.6, 655.6, and 660.6 nm). At each dressing-field
wavelength, two types of spectra are computed. The
spectra labeled 2D were obtained as described above, while the spectra labeled 1D
were computed by restricting the field-free eigenstates in the basis
to those having $J=0$ or $1$.
This restriction in the rotational
quantum number leads to the inhibition of rotational motion, thus
the dissapearance of the LICI.\cite{LICI5} Therefore, all spectra obtained in
the 1D model inherently lack any signatures originating from a LICI.

\begin{figure}[H]
	\begin{centering}
	\includegraphics[width=0.45\textwidth]{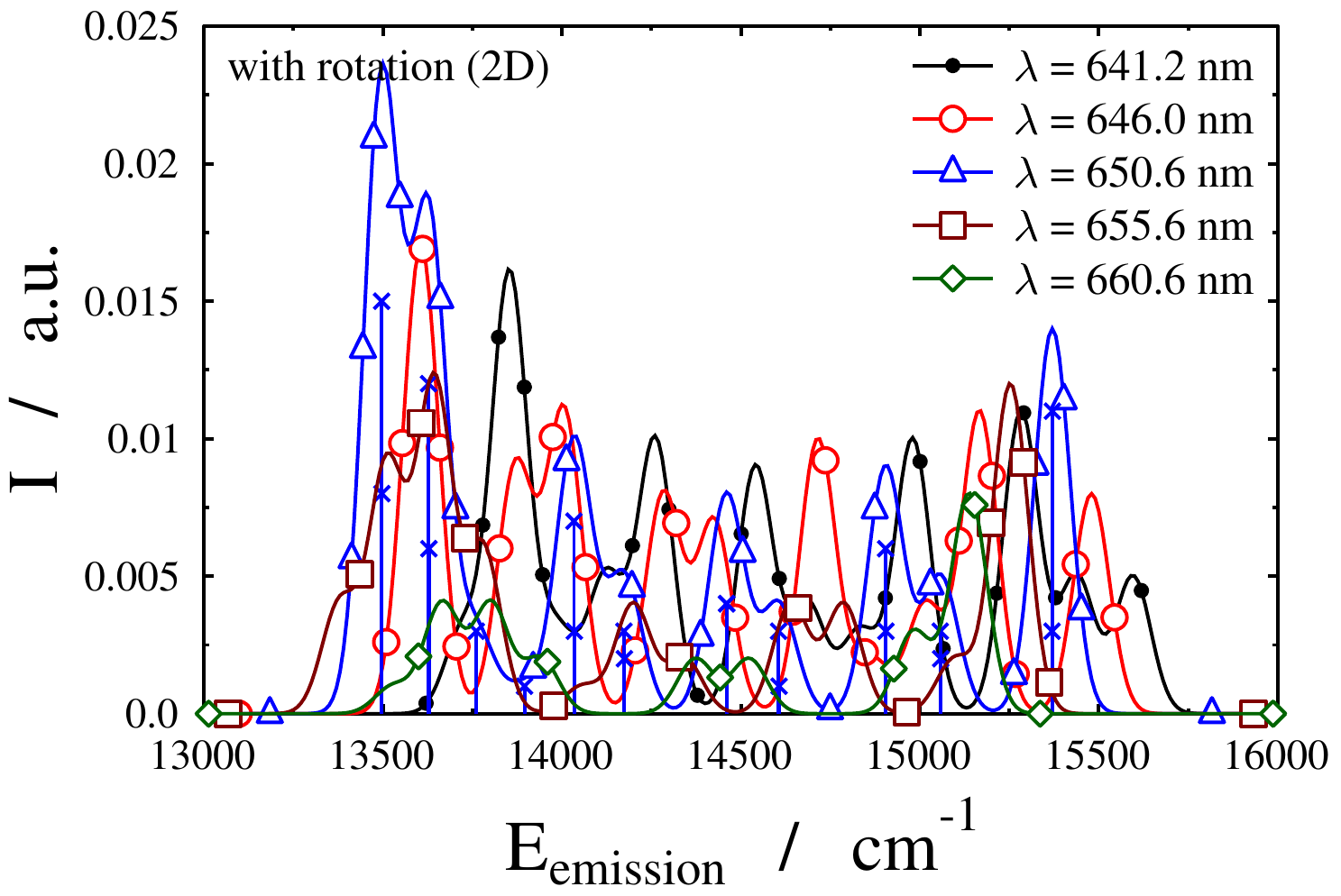} 
	\includegraphics[width=0.45\textwidth]{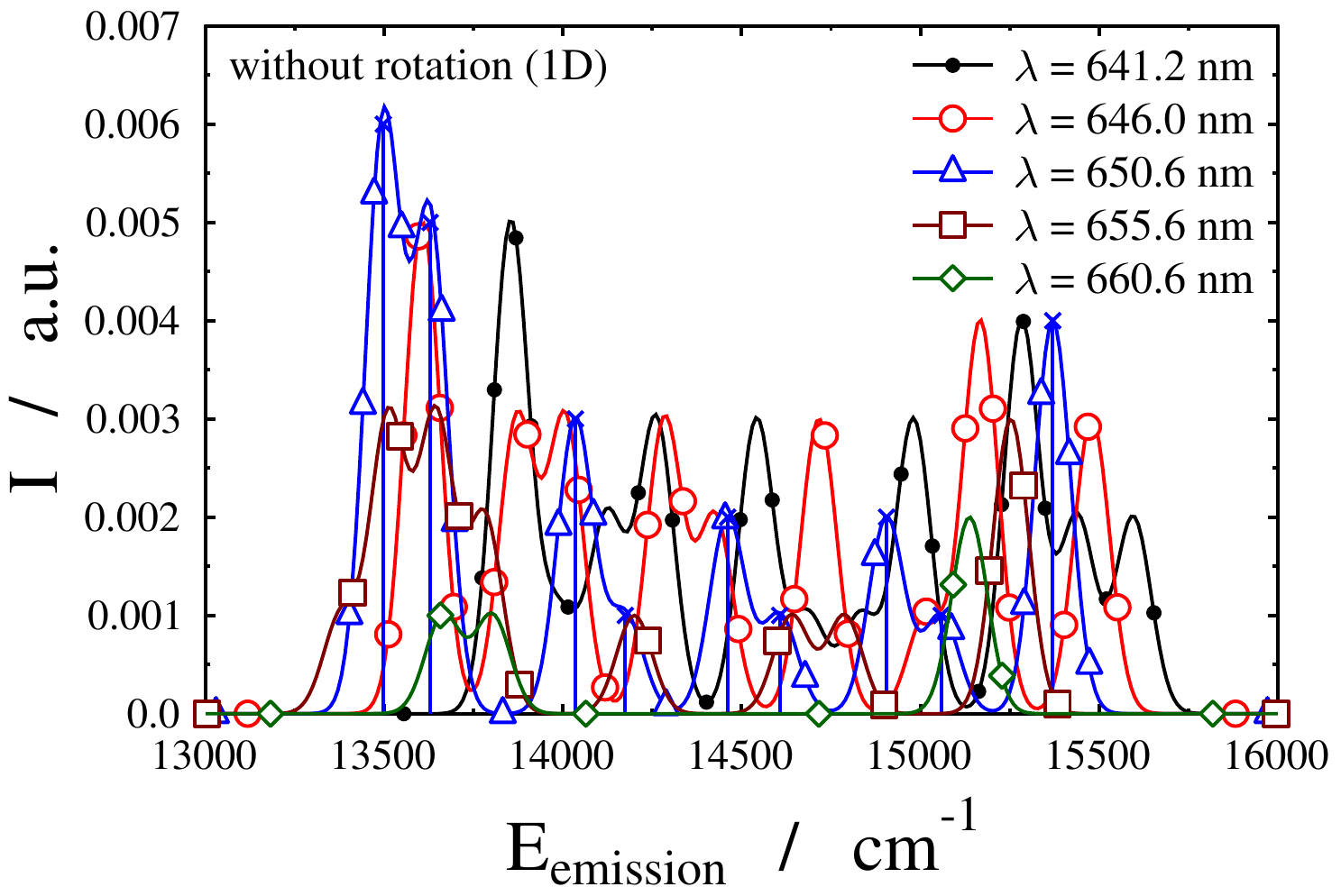} 
	\includegraphics[width=0.45\textwidth]{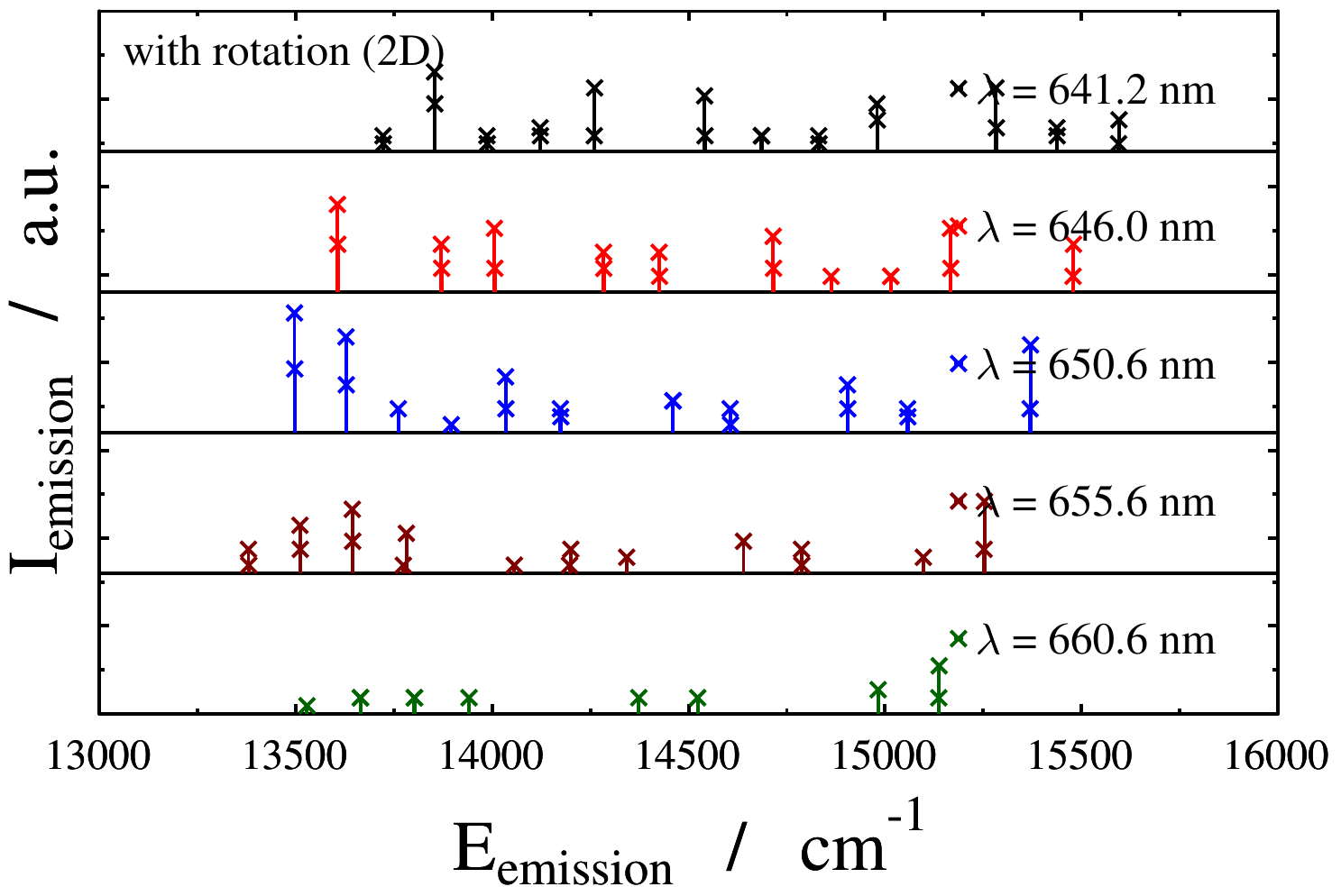}
	\includegraphics[width=0.45\textwidth]{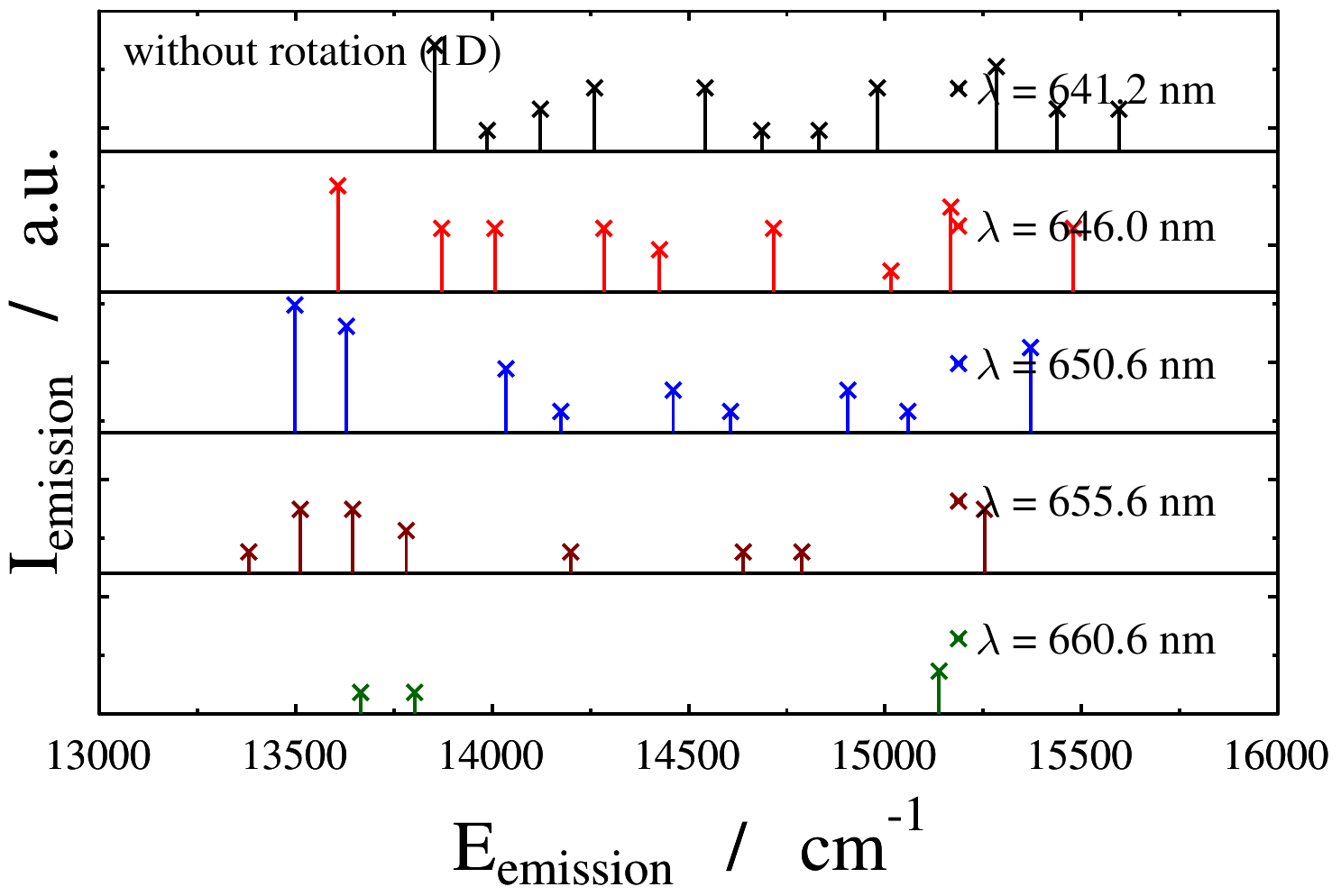} 
	\par\end{centering}

	\caption{
	Stimulated emission spectra of the field-dressed Na$_{2}$ at five non-resonant
dressing-field wavelengths and a dressing-field intensity of $I=10^{8}$ Wcm$^{-2}$, obtained within the 2D (two panels on the lefthand side) or 1D model (two panels on the righthand side), see text.}

	\label{Fig:Fig5} 
\end{figure}

Fig. \ref{Fig:Fig4} demonstrates that the 2D spectra exhibit
significantly more absorption peaks than their 1D counterparts. This is expected, since in the 2D case field-free states
with $J>1$ rotational quantum number are populated and mixed in to
the field-dressed states, which together with the $\Delta J=\pm1$
selection rule for one-photon processes lead to the appearance of
many additional transition peaks. 
In the adiabatic representation of field-dressed PECs, the large number of absorption peaks in the 2D spectra are identified as signatures of either the strong rovibrational coupling arising from the $\theta$-dependent adiabatic PECs or the nonadiabatic couplings near the LICI.
Figure \ref{Fig:Fig4} also demonstrates that the peak intensities in the 2D spectra are more sensitive to the dressing-field wavelength than their 1D counterparts. This is also visible in the stimulated emission peaks, as shown in Fig. \ref{Fig:Fig5}.

Interestingly, variations in the intensity of both the absorption and stimulated emission peaks in the 2D case seem to be most
prominent when the field-dressed state
correlating to $\vert1$ $0$ $0\rangle$ at $F_0\rightarrow0$ has a quasienergy located near the energy position of the LICI. 
In the 1D case, variations of the peak intensities in the vicinity of the diabatic PEC crossing (there is no LICI in the 1D case) seem to
be less pronounced in the stimulated emission spectra and are barely visible in the absorption spectra. 
In classical molecular spectroscopy, the presence of an intrinsic conical intersection and non-adiabatic couplings are known to cause the breakdown of the Frank--Condon principle\cite{spectroscopy_HHRS} and lead to irregular variations in peak intensities, often referred to as \textit{intensity borrowing}.\cite{Cederbaum_multimode,BunkerJensen} 
This leads to the conclusion that the significant intensity variation of the peaks in the 2D spectra near a LICI originates from strong non-adiabatic effects induced by the LICI.

\section{Summary and conclusions}

This study investigates direct signatures of LICIs on the spectra of molecules.
The theoretical framework and the working equations have been formulated for computing one-photon transitions between field-dressed rovibronic states.
The field-dressed states of our showcase molecule, Na$_{2}$, have been computed for a 
The field-dressed spectra feature absorption peaks as well as stimulated emission peaks. The absorption peaks resemble the field-free spectra, while the stimulated emission peaks correspond to transitions not visible in the field-free case.
By investigating the dressing-field wavelength dependence of the field-dressed spectra for both full- and reduced-dimensional simulations, direct signatures of the LICI in the field-dressed spectrum are identified.
These signatures are (1) the appearance of new peaks and the splitting of peaks for both absorption and stimulated emission, and (2) the manifestation of an intensity borrowing effect, \textit{i.e}, an increase in the overall peak intensities when the quasienergy of the initial field-dressed state is in the vicinity of the energy position of the LICI. These signatures of the LICI originate from the strong rovibronic coupling in the field-dressed adiabatic electronic states and the non-adiabatic coupling induced by the LICI. 

\begin{acknowledgement}
This research was supported by the EU-funded Hungarian grant EFOP-3.6.2-16-2017-00005.
The authors acknowledge financial support by the Deutsche Forschungsgemeinschaft
(Project ID CE10/50-3). The authors are grateful to NKFIH for
support (Grant No. PD124623 and K119658) 
\end{acknowledgement}

\providecommand{\latin}[1]{#1}
\providecommand*\mcitethebibliography{\thebibliography}
\csname @ifundefined\endcsname{endmcitethebibliography}
  {\let\endmcitethebibliography\endthebibliography}{}

\end{document}